\documentclass{iopart}
\usepackage{iopams}
\usepackage{graphicx}
\usepackage{setstack}
\def\NL{N_{1}}
\def\NR{N_{2}}
\def\ph{\hat{\phi}}
\def\fh{\hat{f}}
\def\wh{\hat{w}}
\def\Imag{\mathrm{Im}}

\newcommand{\frefs}[1]{figures~\ref{#1}}
\newcommand{\srefs}[1]{sections~\ref{#1}}

\begin{document}

\title{Transport moments and Andreev billiards with tunnel barriers}

\author{Jack Kuipers and Klaus Richter}
\address{Institut f\"ur Theoretische Physik, Universit\"at Regensburg, D-93040
Regensburg, Germany}
\ead{jack.kuipers@ur.de}

\begin{abstract}
Open chaotic systems are expected to possess universal transport statistics and recently there have been many advances in understanding and obtaining expressions for their transport moments.  However when tunnel barriers are added, which represents the situation in more general experimental physical systems much less is known about the behaviour of the moments.  By incorporating tunnel barriers in the recursive semiclassical diagrammatic approach we obtain the moment generating function of the transmission eigenvalues at leading and subleading order.  For reflection quantities quantum mechanical tunneling phases play an essential role and we introduce new structures to deal with them.  This allows us to obtain the moment generating function of the reflection eigenvalues and the Wigner delay times at leading order.  Our semiclassical results are in complementary regimes to the leading order results derived from random matrix theory expanding the range of theoretically known moments.  As a further application we derive to leading order the density of states of Andreev billiards coupled to a superconductor through tunnel barriers.
\end{abstract}


\section{Introduction}\label{intro}

Quantum systems that are chaotic in the classical limit exhibit universal behaviour, for example for the transport through quantum dots \cite{marcusetal92,changetal94}.  For open systems, the transport properties are encoded in the scattering matrix connecting the incoming and outgoing wavefunctions in the leads.  If we imagine a chaotic cavity attached to two scattering leads carrying $\NL$ and $\NR$ channels respectively, with a total of $N=\NL+\NR$, then the scattering matrix separates into four blocks
\begin{equation}
S(E) = \left(\begin{array}{cc}\boldsymbol{r}&\boldsymbol{t}' \\ \boldsymbol{t} & \boldsymbol{r}'\end{array}\right).
\end{equation}
Of particular interest are the transmission eigenvalues and their moments $\Tr \left[\boldsymbol{t}^\dagger \boldsymbol{t}\right]^n$ given in terms of the transmission subblock of the scattering matrix.  The transmission eigenvalues relate to electronic transport through the cavity, like the conductance which is proportional to their first moment \cite{landauer57,buttiker86,landauer88} and the power of the shot noise which is related to their second.

To obtain a handle on the transport moments, one can employ the semiclassical approximation for the elements of the scattering matrix \cite{miller75,richter00} 
\begin{equation} 
  \label{scatmateqn}
  S_{oi}(E) \approx \sqrt{\frac{\mu}{N}}\sum_{\gamma (i \to o)}
  A_{\gamma}(E)\rme^{\frac{\rmi}{\hbar}S_{\gamma}(E)} ,
\end{equation}
in terms of all the classical scattering trajectories $\gamma$ that connect the corresponding channels.  They contribute with their stability amplitude $A_{\gamma}$ and a phase involving their action $S_{\gamma}$ while $\mu$ is the escape rate of the corresponding classical system.

The first moment of the transmission eigenvalues then follows from a sum over pairs of trajectories which start in the same channel in one lead and end in the same channel in the other
\begin{equation} 
  \label{transfirstmoment}
  \Tr \left[\boldsymbol{t}^\dagger \boldsymbol{t}\right] \approx \frac{\mu}{N}\sum_{i=1}^{\NL}\sum_{o=1}^{\NR}\sum_{\substack{\gamma (i \to o) \cr \gamma' (i \to o)}}  A_{\gamma}A^{*}_{\gamma'}\rme^{\frac{\rmi}{\hbar}\left(S_{\gamma}-S_{\gamma'}\right)} .
\end{equation}
This fluctuates as the energy is varied and as we are particularly interested in the statistics of the transmission eigenvalues, we average over a range of energies.  The averaging over the phase difference picks out pairs of trajectories with highly similar actions and indeed when the trajectories are identical ($\gamma=\gamma'$), one recovers the contribution $N_1N_2/N$ \cite{bjs93a,bjs93b} which is the leading order term of an expansion in $N^{-1}$ of the full result.  The next order term was discovered to follow from trajectories that come close to themselves in an `encounter' as they travel through the cavity \cite{rs02}.  A partner trajectory can then be found which is nearly identical, but which traverses the encounter differently. Together these pairs provide the weak localisation correction $-N_1N_2/N^2$ for systems with time reversal symmetry.  Higher order corrections came from more complicated pairs of trajectories which are nearly identical in long stretches called `links' while differing in small encounter regions and which could be summed to give the complete result for the first moment \cite{heusleretal06}.

\begin{figure}
  \centering
  \includegraphics[width=0.73\textwidth]{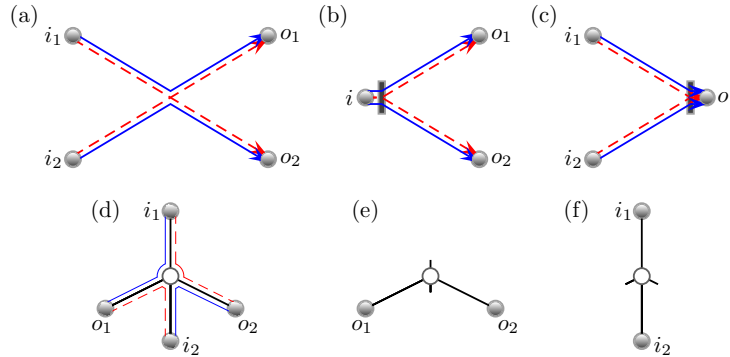}
  \caption{The trajectory quadruplet which meets at a central encounter in (a) contributes to the leading order term of the second moment.  The quadruplet can be redrawn as the tree in (d) where the paths around the trees recreate the trajectories in (a).  By moving the encounter to the incoming lead, with $i_1=i_2$, we obtain the possible trajectory configuration in (b), here shown with the trajectories tunneling through the barrier.  Removing the paths on the left of the encounter can be represented by leaving stubs in the tree diagram as in (e).  Moving the encounter to the outgoing lead provides the structure in (c) or the tree diagram in (f).}
  \label{secondmoment}
\end{figure}

The semiclassical treatment of the first two moments and other correlations functions \cite{heusleretal06,braunetal06,mulleretal07} led to diagrammatic rules where the contribution of any diagram could be read off from its structure.  In particular, each link provides a factor of $N^{-1}$, each encounter a factor of $-N$ and each channel in the first or second lead a factor of $\NL$ or $\NR$.  The main leading order diagram for the second moment of the transmission eigenvalues, which is pictured in \fref{secondmoment}(a) then provides a contribution of $-\NL^2\NR^2/N^3$.  From this diagram it is further possible to shift the encounter to the left until it enters the incoming lead as in \fref{secondmoment}(b).  We note that this picture of encounters moving into leads actually derives from the semiclassical treatment of Ehrenfest time effects \cite{jw06,br06,wj06} where the remnant of the encounter provides an Ehrenfest time dependent factor which ensures the unitarity of the transport.  However, when the Ehrenfest time is small compared to the dwell time (the typical time that trajectories spend inside the cavity), the semiclassical contribution is as if the encounter is removed entirely and the channels in the lead coincide.  The configuration in \fref{secondmoment}(b) then has $i_1=i_2$ and provides a contribution of $\NL\NR^2/N^2$.  Likewise, the corresponding case of moving the encounter into the outgoing lead in \fref{secondmoment}(c) provides the contribution $\NL^2\NR/N^2$.  These three cases are the only possibilities at leading order, giving a total of $N\xi(1-\xi)$ with $\xi=\NL\NR/N^2$.

With the diagrammatic rules, the problem of calculating the moments reduces to that of finding all the possible diagrams, which was performed for the first two moments \cite{heusleretal06,braunetal06,mulleretal07} through the connection to correlated periodic orbits which contribute to spectral statistics of closed systems \cite{sr01,mulleretal04,mulleretal05}.  For higher moments instead, the leading order diagrams can be represented as trees \cite{bhn08}.  For example, the trajectories in \fref{secondmoment}(a) become the tree in \fref{secondmoment}(d) while moving encounters into the leads corresponds to removing alternating links around the encounter node as in \frefs{secondmoment}(e) and (f).  Since trees can be generated recursively by cutting them at nodes into sets of smaller trees, all the moments of the transmission eigenvalues \cite{bhn08} and the moments of the delay times \cite{bk10} could be obtained at leading order in inverse channel number.  This approach could also be adapted to include energy dependence and obtain the leading order behaviour of the density of states of Andreev billiards \cite{kuipersetal10,kuipersetal11}.  Graphical recursions could then be utilised to obtain all moments of these various quantities at the next few orders in inverse channel number \cite{bk11}.   

Exploring the combinatorial interpretation of the semiclassical diagrams, it was recently shown \cite{novaes12,bk12,novaes13} that they always combine to give exactly the same results as those derived from random matrix theory (RMT) where the scattering matrix is modelled as having random elements and belonging to one of the circular ensembles \cite{beenakker97}.  
In fact the semiclassical diagrams can be related to factorisations of permutations, and one combinatorial interpretation for systems without time reversal symmetry (corresponding to the unitary ensemble) was derived from the periodic orbits of closed systems \cite{novaes12,novaes13}, while the other intepretation for all three classical symmetry classes reduces to primitive factorisations \cite{bk12}.

This equivalence between semiclassics and RMT for transport was originally put forward in the late 80's \cite{bs88,bs90} and RMT was first used to calculate the conductance and its variance \cite{bm94,jpb94} with a diagrammatic approach later providing several leading and subleading order results \cite{bb96}.  However, RMT also provides the probability distribution of the transmission eigenvalues \cite{beenakker97} and their moments are given by integrals related to the Selberg integral.  This connection more recently allowed the shot noise power \cite{ss06} and then various fourth moments to be calculated \cite{ssw08} and brought a lot of interest to the evaluation of these integrals and the corresponding transport moments.  Various techniques were developed to obtain all the moments of the transmission eigenvalues \cite{vv08,novaes08,ms11,lv11} and of the conductance and shot noise \cite{novaes08,ok08,ok09,kss09,ms13} as well as of the Wigner delay times and time delay \cite{ms11,ms13}.  Interestingly, the different techniques tend to lead to different expressions for the moments which are not so obviously related to each other even though they must be equivalent.  Similarly, the RMT and semiclassical moments must agree \cite{novaes12,bk12} but the resulting formulae are different enough to obscure the equivalence.  However, asymptotic analyses \cite{ms12} of the particular expressions for the moments obtained in \cite{ms11} have managed to recover the semiclassically calculated moment generating functions at the first few orders in inverse channel number \cite{bk11}.

The above discussion was for the particular case where the leads couple perfectly to the cavity, but for the more general and important case when this coupling becomes imperfect much less is known.  Imperfect coupling often occurs in physical experiments in systems from microwave billiards to quantum dots and to make the theoretical treatment above of wider practical use we need to expand the framework to include the coupling.  For this we add a thin potential wall or tunnel barrier at the end of the leads so that any incoming (or outgoing) particle is separated into transmitted and reflected parts, which was originally treated semiclassically in \cite{whitney07}.  Here we show how this can be incorporated into the current graphical semiclassical framework \cite{bk11}.

On the RMT side the probability distribution of the transmission eigenvalues is currently unknown, with the state of the art being a cavity with one perfect and one imperfect lead \cite{vk12}.  The moments are likewise generally unknown apart from at leading order when the leads are identical \cite{bb96}.  For the delay times, the probability distribution is known \cite{sss01} but the moments have yet to be evaluated.  In general only the weak localisation correction to the conductance and the universal conductance fluctuations \cite{bb96} along with the subleading contribution to the shot noise power \cite{rbm08} have been derived diagrammatically.  But these have also be obtained semiclassically by considering the diagrams explicitly \cite{whitney07,kuipers09,waltneretal12} a process we now show how to perform implicitly.  

Our paper is organised as follows: In \sref{semitunnels} we explain how tunnel barriers lead to modifications in the number of semiclassical diagrams and on the level of their individual contributions.  In \srefs{trans} and \ref{refl} we derive, to leading order, the moments of the transmission and reflection eigenvalues.  \Sref{wigner} is devoted to the moments of the Wigner delay times while in \sref{andreev} we consider Andreev billiards with tunnel coupled superconducting leads as an important application.  Finally in \ref{subleading} we work out the first subleading order terms for the transmission eigenvalues.

\section{Semiclassics with tunnel barriers} \label{semitunnels}

Introducing tunnel barriers leads to two main changes in the semiclassical diagrammatic treatment of transport.  The first is that the contributions of the individual parts of the diagrams changes.  The tunnel barriers can be treated probabilistically in the semiclassical limit so that trajectories have a certain probability to pass through each time they hit the barriers in the leads \cite{whitney07}.  In general this probability $p_i$ can be different for each channel $i$.  For an $l$-encounter involving $l$ trajectory stretches of length $t$ which are close together, if the encounter hits channel $i$ in the leads, the joint survival probability of all the stretches is $(1-p_i)^l$.  Over time, the survival probability of the encounter is $\rme^{-\mu_l t}$ with an escape rate of
\begin{equation}
\mu_l = \frac{\mu}{N} \sum_{i=1}^{N}1-(1-p_i)^l ,
\end{equation}
where $\mu$ is the escape rate of the same system without tunnel barriers in the lead.
The diagrammatic rules for the contributions of semiclassical diagrams then become \cite{whitney07}:
\begin{itemize}
 \item Each link provides the factor $y=\left(\sum_{i=1}^{N} p_i \right)^{-1}$
 \item Each $l$-encounter the factor $-\sum_{i=1}^{N} 1-(1-p_i)^l$
 \item Each channel sum provides the factor $\sum_{i=1}^{\NL} p_i^j(1-p_i)^k$ or $\sum_{i=\NL+1}^{N} p_i^j(1-p_i)^k$ where $j$ counts the number of trajectory pairs passing through the same channel and $k$ the number of pairs reflected.
\end{itemize}

For example for the moments of the transmission eigenvalues, a diagonal pair starting in a channel in the incoming lead and travelling together to a channel in the outgoing lead give the leading order term of the first moment
\begin{equation}
\label{leadingtransfirst}
\left \langle \Tr \left[\boldsymbol{t}^\dagger \boldsymbol{t}\right] \right \rangle= \frac{\sum_{i=1}^{\NL} p_i\sum_{i=\NL+1}^{N} p_i}{\sum_{i=1}^{N} p_i} + O(1) .
\end{equation}
With the same tunneling probability $p$ in each channel, this simplifies to $pN\xi$ with $\xi=\NL\NR/N^2$ while the standard second moment diagram of \fref{secondmoment}(a) provides
\begin{equation}
-\left(1-(1-p)^2\right)N\frac{p^4\NL^2\NR^2}{p^4N^4} = p(p-2)N\xi^2 .
\end{equation}
Moving the encounter fully into the leads as in \frefs{secondmoment}(b) and (c) gives an additional
\begin{equation}
\frac{p^4\NL\NR^2}{p^2N^2}+\frac{p^4\NL^2\NR}{p^2N^2} = p^2N\xi .
\end{equation}

\begin{figure}
  \centering
  \includegraphics[width=0.8\textwidth]{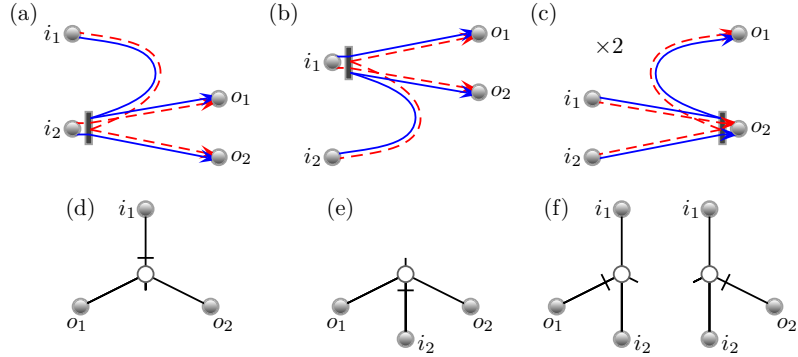}
  \caption{With tunnel barriers, trajectory stretches can additionally be reflected at the tunnel barriers.  When the encounter from \fref{secondmoment}(a) now moves into the lead, one trajectory stretch can tunnel through while the other is reflected as in (a)--(c).  In the tree representation there are four possibilities of removing one link (represented as a stub) and reflecting the other (represented by the perpendicular bar) as in diagrams (d)--(f).}
  \label{secondmomenttunnel}
\end{figure}

The second main change with tunnel barriers is that many additional diagrams become possible \cite{whitney07,kuipers09}.  Encounters can now partially enter the leads as trajectories can be reflected back into the cavity and are no longer forced to leave the system.  Staying with the simpler example of the moments of the transmission eigenvalues, from the diagram in \fref{secondmoment}(a), we could move the encounter into either lead, let one trajectory pair stretch pass through the tunnel barrier and one be reflected as in \frefs{secondmomenttunnel}(a)--(c).  From the channel sum of the encounter touching the lead, we obtain a factor of $p(1-p)N_1$ or $p(1-p)N_2$ and an additional contribution of
\begin{equation}
2p(1-p)\NL\frac{p^3\NL\NR^2}{p^3N^3}+2p(1-p)\NR\frac{p^3\NL^2\NR}{p^3N^3}= 4p(1-p)N\xi^2 ,
\end{equation}
where the factor 2 comes from the two possibilities of which trajectory to let pass through the barrier.  The total result for the leading order term of the second moment of the transmission eigenvalues becomes
\begin{equation}
\label{leadingtranssecond}
\left \langle \Tr \left[\boldsymbol{t}^\dagger \boldsymbol{t}\right]^2 \right \rangle= pN\xi(p+2\xi-3p\xi) + O(1) .
\end{equation}

Our aim now is to systematically generate these additional possible diagrams, with their semiclassical contributions, for higher moments.

\section{Moments of the transmission eigenvalues} \label{trans}

As the moments of the transmission eigenvalues involve trajectories starting in one lead and ending in the other, the possibilities for moving encounters partially into the leads are somewhat limited.  They can therefore be treated relatively easily by extending the treatment without tunnel barriers and here we mainly highlight the changes needed to incorporate them.  At leading order in inverse channel number, the underlying semiclassical diagrams can be redrawn as trees \cite{bhn08} as going from the top lines in \fref{secondmoment} and \fref{secondmomenttunnel} to the bottom lines.  This framework has been further developed \cite{bk10,kuipersetal10} and detailed in \cite{kuipersetal11}.  Graphical recursions can be used to go beyond leading order and we build on that formalism here using generating functions as defined in \cite{bk11}.  In particular we will obtain an expansion for the moment generating function
\begin{equation}
T(s) = \sum_{n=1}^{\infty}s^n \left \langle \Tr \left[\boldsymbol{t}^\dagger \boldsymbol{t}\right]^n \right \rangle = NT_0 + T_1 + \ldots
\end{equation}

\subsection{Tree recursions}

\begin{figure}
  \centering
  \includegraphics[width=0.57\textwidth]{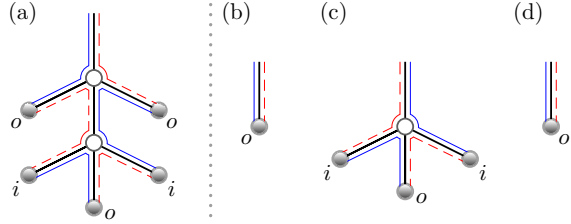}
  \caption{For the tree recursions we start with unrooted trees which come in two types, those with an excess of outgoing channels as in (a) and (b) which belong to $\phi$ and those with an excess of incoming channels as in (c) which belong to $\ph$.  By combining the top of the three trees in (b), (c) and (d) into a new encounter and adding a new link, we can create the tree in (a).}
  \label{treerec}
\end{figure}

To generate the leading order semiclassical diagrams, we start from the related tree diagrams where the encounters become vertices of even degree ($\geq4$), the links become edges and the incoming and outgoing channels become leaves or vertices of degree 1.  The boundary walk of the tree allows us to recover how the semiclassical trajectories are arranged.  For example, the trajectories in \fref{secondmoment}(a) become the boundary walk of the tree in \fref{secondmoment}(d) and vice versa.  This also means that the incoming and outgoing channels must alternate around the tree itself.  To generate the trees recursively we start by not rooting them in any channel so that these intermediate trees have an odd number of leaves and therefore come in two types.  One type has an excess of outgoing channels so that, if it had a root, the root would be an incoming channel and we say that this type of tree starts from an incoming direction.  For example, the tree in \fref{treerec}(a) is of this type.  The other type has an excess of incoming channels and starts from an outgoing direction as in \fref{treerec}(c).  

We then let $\phi$ and $\ph$ be generating functions which count all trees (including their semiclassical contributions) starting inside the system from an incoming or outgoing direction respectively and not rooted in any channel.  The function $\phi$ includes the trees in \frefs{treerec}(a) and (b) as well as that in \fref{secondmoment}(d) with the channel $i_1$ at the top removed.
As the $n$th moment involves $2n$ semiclassical trajectories with $n$ incoming and $n$ outgoing channels, by including a factor $r$ with each channel we can track which moment each tree diagram contributes to.  The variable $r$ then becomes the generating variable so that later the coefficient of $r^{2n}$ will give the $n$th moment.

We can generate trees recursively since the trees start with a link which connects to an encounter of arbitrary size $l$, while alternating below it are $l$ further trees of the same type as the starting tree and $l-1$ trees of the opposite type.   An example for $l=2$ can be seen in \fref{treerec}(a).  Similarly, bringing trees together in this way allows us to create all larger trees.  Trees can be brought together only if they all start inside the system, which is why we have this restriction on $\phi$ and $\ph$.

Without tunnel barriers, for $\phi$ we could further remove all the $l$ trees of $\phi$ type below the encounter and move this encounter into the outgoing lead, as in \fref{secondmoment}(f).  We include the contribution of such diagrams (with the top channel removed again) in $\phi$, but since the start point of $\phi$ is supposed to be inside the system, the diagram in \fref{secondmoment}(e) is not included.  Such cases will be accounted for later.

With tunnel barriers the new possibilities for the encounter are illustrated for the second moment in \fref{secondmomenttunnel}. In general, touching the outgoing lead involves removing any $k$ of the $l$ trees and allowing the other $l-k$ to be reflected back into the cavity, as in \fref{secondmomenttunnel}(f).  The case $k=0$ just means that the entire encounter is reflected at the tunnel barrier and is already included in the semiclassical contribution of the encounter.  The encounter could also move to the incoming lead with the first link from the starting point to the encounter reflected and $k$ of the $l-1$ trees of type $\ph$ removed to tunnel straight into the lead instead.  This possibility is illustrated in \fref{secondmomenttunnel}(d), while a diagram like that in \fref{secondmomenttunnel}(e) is not included in $\phi$.  Again the case $k=0$ is already included elsewhere and we obtain the recursion relation
\begin{eqnarray}
\label{phirecur}
\phi &=& yr\sum_{i=\NL+1}^{N} p_i - y\sum_{l=2}^\infty \sum_{i=1}^{N} \left(1-(1-p_i)^l\right)\phi^{l}\ph^{l-1} \nonumber \\
& & {} +y\sum_{l=2}\sum_{i=\NL+1}^{N}\ph^{l-1}\sum_{k=1}^l \left(\begin{array}{c}l \\ k \end{array} \right)\phi^{l-k}(1-p_i)^{l-k}p_i^kr^k \nonumber \\
\fl & & {} +y\sum_{l=2}\sum_{i=1}^{\NL}\phi^{l}\sum_{k=1}^{l-1} \left(\begin{array}{c}l-1 \\ k \end{array} \right)\ph^{l-1-k}(1-p_i)^{l-k}p_i^kr^k .
\end{eqnarray}
The first line starts with the contribution of the smallest tree which is just a link that tunnels through into the outgoing lead as in \fref{treerec}(b).  In the first line we then add all the trees whose top encounter does not move into the lead.  The remaining two lines count the possibilities that the encounter moves into the outgoing or incoming lead and $k$ of the stretches of the encounter tunnel straight through the lead (with the remaining stretches being reflected back into the cavity).  Since the powers of $r$ count the order of the diagram, a factor $r$ is included with each tree which is removed along with a corresponding factor of $p$ to record that those links have tunnelled through the barrier.

In the recursion relation we can perform the sums over $k$.  Since the $(1-p)^l$ in the top line of \eref{phirecur} corresponds to the $k=0$ terms of both sums, we can simplify to
\begin{eqnarray}
\label{phirecursum}
\fl \phi\sum_{i=1}^{N} p_i &=& r\sum_{i=\NL+1}^{N} p_i - N\sum_{l=2}^\infty \phi^{l}\ph^{l-1} 
+\sum_{l=2}\sum_{i=\NL+1}^{N}\ph^{l-1}\left(rp_i+\phi(1-p_i)\right)^{l}\nonumber \\
\fl & & {} +\sum_{l=2}\sum_{i=1}^{\NL}\phi^{l}(1-p_i)\left(rp_i+\ph(1-p_i)\right)^{l-1} ,
\end{eqnarray}
where we further divided both sides by $y$.  Neatly, the first two terms can be rearranged into the $l=1$ terms of the three sums over $l$, so that the recursion reduces to
\begin{equation}
\label{phirecursummed}
\fl \frac{N\phi}{1-\phi\ph} = \sum_{i=\NL+1}^{N}\frac{rp_i+\phi(1-p_i)}{1-rp_i\ph-\phi\ph(1-p_i)}+\sum_{i=1}^{\NL}\frac{\phi(1-p_i)}{1-rp_i\phi-\phi\ph(1-p_i)} .
\end{equation}
For $\ph$ we likewise have
\begin{equation}
\label{phihatrecursummed}
\fl \frac{N\ph}{1-\phi\ph} = \sum_{i=1}^{\NL}\frac{rp_i+\ph(1-p_i)}{1-rp_i\phi-\phi\ph(1-p_i)}+\sum_{i=\NL+1}^{N}\frac{\ph(1-p_i)}{1-rp_i\ph-\phi\ph(1-p_i)} .
\end{equation}

\subsection{Leading order moments}

To obtain the leading order diagrams, we now root the top of $\phi$ in an incoming channel.  This also allows us to remove the top link and tunnel straight into an encounter in the incoming lead, so we can now add diagrams like \fref{secondmoment}(e) and \fref{secondmomenttunnel}(e).  Of the remaining $l-1$ alternating $\ph$ trees emanating from the encounter, any $k$ can also move straight into the lead.  The generating function $T_0$, which counts all the leading order diagrams with their contributions and divided by $N$, is then given by
\begin{equation}
\label{Phieqn1}
\fl NT_0 = \sum_{i=1}^{\NL}rp_i\phi + \sum_{i=1}^{\NL}p_ir\sum_{l=2}^{\infty}\phi^{l}\sum_{k=0}^{l-1} \left(\begin{array}{c}l-1 \\ k \end{array} \right)\ph^{l-1-k}(1-p_i)^{l-1-k}p_i^kr^k ,
\end{equation}
Performing the sums gives
\begin{equation}
\label{Phieqngeneral}
NT_0 = \sum_{i=1}^{\NL}\frac{rp_i\phi}{1-rp_i\phi-\phi\ph(1-p_i)} .
\end{equation}

\subsection{Fixed tunneling probabilities in each lead}

With the formulae in \eref{phirecursummed},~\eref{phihatrecursummed} and~\eref{Phieqngeneral} we have a formal generating function for the leading order terms of all moments and we can expand $\phi$, $\ph$ and then $T_0$ in powers of $r$.  However, with the sums over the different tunneling probabilities in each channel, it is difficult to manipulate the expressions further.  To proceed we can make the simplifying assumption that the channels in each lead have the same tunneling probability of $p_1$ for the incoming lead and $p_2$ for the outgoing lead.  Equations~\eref{phirecursummed} and~\eref{phihatrecursummed} lead to quartic equations for $\phi$ and $\ph$ which in turn lead to the following quartic equation for $T_0$:
\begin{eqnarray}
\label{Phiquarticdiffleads}
 (s-1)\left(s^2p_1^2p_2^2+2sp_1p_2(2-p_1-p_2)+(p_1-p_2)^2\right)T_0^4 \nonumber \\
 + 2s(s-1)p_1p_2\left(sp_1p_2+2-p_1-p_2\right)T_0^3 \nonumber \\
 + s\left[s(s-1)p_1^2p_2^2+p_1^2(p_2-1)\zeta_1+p_2^2(p_1-1)\zeta_2 \right. \nonumber \\
 \qquad \left. +\xi\left(2sp_1p_2(2-p_1-p_2)+s(2s-1)p_1^2p_2^2+(p_1-p_2)^2\right)\right]T_0^2 \nonumber \\
 +s^2p_1^2p_2^2\xi(2s-1)T_0+s^3p_1^2p_2^2\xi^2 = 0 ,
\end{eqnarray}
where $\zeta_1=\NL/N$, $\zeta_2=\NR/N$ and $r^2=s$ is the generating variable for the moments since the $n$th moment involves trees with $2n$ leaves.  Expanding $T_0$ in powers of $s$ we need to choose the correct value for the first moment, which is $p_1p_2\xi/(p_1\zeta_1+p_2\zeta_2)$ from \eref{leadingtransfirst}.

\subsection{One perfect lead}

Equation~\eref{Phiquarticdiffleads} allows us for example to consider the situation where only one of the leads has a tunnel barrier by setting $p_1$ or $p_2$ to 1, as in the recently considered situation where the probability distribution of the reflection eigenvalues was obtained from RMT for systems without time reversal symmetry \cite{vk12} although with different tunneling probabilities in the channels in the remaining lead.  With equal probabilities instead and $p_2=1$, we actually obtain a cubic equation semiclassically
\begin{eqnarray}
\label{Phicubiconelead}
(s-1)\left(1+p_1(s-1)\right)T_0^3 + s(s-1)p_1(1+\zeta_2)T_0^2 \nonumber \\
+ s\left(\xi+p_1\xi(s-1) + \zeta_2(sp_1-1)\right)T_0 +s^2p_1\xi\zeta_2 = 0 ,
\end{eqnarray}
which can also be obtained from \eref{Phiquarticdiffleads} through using $\xi=\zeta_1\zeta_2$ and $\zeta_1=1-\zeta_2$.  The generating function in \eref{Phicubiconelead} should then match the leading order moments found by integrating the corresponding transmission probability distribution to that in \cite{vk12}. 

\subsection{Equal tunneling probabilities}

To proceed further we instead set all of the tunneling probabilities equal to $p=p_1=p_2$.  The quartic equation for $\phi$ now has the expansion 
\begin{eqnarray}
\label{phiexpansion}
\fl \phi &=&  \zeta_2 r + \xi\left(p+\zeta_2(1-2p)\right)r^3 + \nonumber \\
\fl  & & {} + \xi\left(p^2+p\zeta_2(1-2p)+p\xi(3-5p)+2\zeta_2\xi[1-5p(1-p)] \right)r^5 \ldots
\end{eqnarray}
while for $\ph$ we swap $\zeta_2$ and $\zeta_1$.  The second term in \eref{phiexpansion} corresponds to the sum of the diagrams in \frefs{secondmoment}(d) and (f) and \frefs{secondmomenttunnel}(d) and (f).  The quartic for $T_0$ reduces to
\begin{eqnarray}
\label{Phiquartic}
(s-1)\left(sp^2+4(1-p)\right)T_0^4 + 2(s-1)\left(sp^2+2(1-p)\right)T_0^3 \nonumber \\
+ \left[s(s-1)p^2+p-1+s\xi\left(4(1-p)+(2s-1)p^2\right)\right]T_0^2 \nonumber \\
+s(2s-1)p^2\xi T_0+s^2p^2\xi^2 = 0 ,
\end{eqnarray}
and again for the expansion in powers of $s$ we need to pick the correct value of $p\xi$ for the first moment (since we divided $T_0$ by $N$).  The first few terms in the expansion are
\begin{eqnarray}
\label{T0expansion}
T_0 & = & p\xi s + p\xi\left(p+2\xi-3p\xi\right)s^2 \nonumber \\
& & {} +p\xi\left(p^2+2p(3-4p)\xi+(6-21p+17p^2)\xi^2\right)s^3 +\ldots 
\end{eqnarray}
with the second term corresponding to the second moment calculated explicitly in \sref{semitunnels}.

\subsection{Equal leads}

As a further simplification, we can also have an equal number of channels in each lead so that $\zeta_1=\zeta_2=1/2$.  Due to the symmetry we have $\phi=\ph$, so \eref{phirecursum} reduces to
\begin{equation}
\label{phiequalleads}
\frac{2\phi}{1-\phi^2} = \frac{rp+2\phi(1-p)}{1-rp\phi-\phi^2(1-p)}, \qquad r\phi^2-2\phi+r = 0 ,
\end{equation}
where $\phi$ is actually given by the same quadratic as when there are no tunnel barriers in the leads.  The moment generating function then satisfies the quadratic equation
\begin{equation}
\label{Phisimpquadratic}
4(s-1)\left(sp^2+4(1-p)\right)T_0^2 + 4(s-1)sp^2T_0 + s^2p^2 = 0 ,
\end{equation}
which is also a factor of \eref{Phiquartic} when $\zeta_1=\zeta_2=1/2$ or $\xi=1/4$.  The moment generating function can then be given explicitly as
\begin{equation}
\label{Phiquadraticmgf}
T_0=\frac{sp\left(p(1-s)+(p-2)\sqrt{1-s}\right)}{2(s-1)\left(sp^2+4(1-p)\right)} ,
\end{equation}
where we chose the solution of \eref{Phisimpquadratic} which matches the first moment $p/4$.

\subsection{Summary of different results}

We summarize the restrictions for the results in the different cases above and detail which generating function is appropriate for which situation in Table~\ref{transgenfuncts}.

\Table{\label{transgenfuncts}Leading order generating functions for the moments of the transmission eigenvalues for different restrictions on the tunneling probability and number of channels in each lead.}
\begin{tabular}{cccc}
\br
Tunneling probability & Tunneling probability & Equal number & Equation \\
in the first lead & in second lead & of channels & for $T_0$ \\
\mr
$p_1$ & $p_2$ & no & \eref{Phiquarticdiffleads} \\
$p_1$ & 1 & no & \eref{Phicubiconelead} \\
$p$ & $p$ & no & \eref{Phiquartic} \\
$p$ & $p$ & yes & \eref{Phiquadraticmgf} \\
\br
\end{tabular}
\endTable

\subsection{Comparison with RMT}

The RMT result for the leading order probability distribution of the transmission eigenvalues was calculated in \cite{bb96}.  However, a final result could be obtained if they assumed the two leads were identical.  The channels could have different individual tunneling probabilities though, as long as the probabilities are matched in the other lead.  The result \cite{bb96} was
\begin{equation} \label{RMTprobdensity}
P(Z)= \sum_{i=1}^{N_1}\frac{p_i(2-p_i)}{\pi(p_i^2-4p_iZ+4Z)\sqrt{Z(1-Z)}} .
\end{equation}
Semiclassically instead it is simple to have different sized leads, but with a constant tunneling probability in each.  To compare the different results we can look at the common results for the simplest case of identical leads with a single tunneling probability.  From the moment generating function in \eref{Phiquadraticmgf}, including the 0th moment as 1, we can perform the Hilbert transformation to get the probability density
\begin{equation}
\label{semiclassicalprobdensity}
\fl 1+T_0=\int_{0}^{1}\frac{P(Z)}{N} \frac{1}{1-sZ} \rmd Z, \qquad  \frac{P(Z)}{N}= \frac{p(2-p)}{2\pi(p^2-4pZ+4Z)\sqrt{Z(1-Z)}} ,
\end{equation}
which is exactly \eref{RMTprobdensity} with identical tunneling probabilities.  That the distribution in \eref{RMTprobdensity} for identical leads is a sum over the channels means that the moment generating function would likewise be a sum over terms like in \eref{Phiquadraticmgf} again with different $p_i$.  Semiclassically, with identical leads we have $\phi=\ph$ through symmetry, but this result still cannot easily be seen from \eref{phirecursummed} and \eref{Phieqngeneral}.  Instead we have access to the complementary regime of different leads with equal tunneling probabilities.

We can proceed to higher orders in the inverse number of channels (see \cite{rs02}) by similarly modifying the approach \cite{bk11} without tunnel barriers.  For example, we present the calculation of $T_1$ in \ref{subleading}.  However, the more interesting case occurs when we consider reflection quantities.

\section{Moments of the reflection eigenvalues} \label{refl}

\begin{figure}
  \centering
  \includegraphics[width=0.6\textwidth]{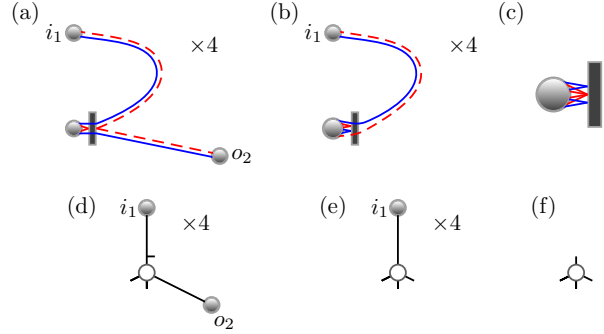}
  \caption{For reflection quantities incoming and outgoing channels can coincide so that with tunnel barriers new possibilities arise for encounter stretches to tunnel into the lead.  Allowing the stretch to channel $o_2$ from \fref{secondmomenttunnel}(a) to also tunnel into the lead so that channels $o_1=i_2$ give the possible diagram in (a).  In the graphical representation in (d) this corresponds to removing adjacent links around the encounter node, which can be performed in 4 ways.  We can remove an additional link as in (b) and (e), if $o_2=o_1=i_2$ where now a pair of trajectories are directly reflected at the tunnel barrier, and finally have the situation in (c) and (f) where none of the trajectories ever enter the system.}
  \label{secondmomentreflection}
\end{figure}

When we consider the moments of the reflection eigenvalues, or their generating function
\begin{equation}
R(s) = \sum_{n=1}^{\infty}s^n \left\langle \Tr\left[\boldsymbol{r}^\dagger \boldsymbol{r}\right]^n \right\rangle = NR_0 + R_1 + \ldots
\end{equation}
we now have a semiclassical approximation where the trajectories all start and end in the same lead.  With tunnel barriers, this allows even more diagrammatic possibilities when moving encounters into the lead, and we may also have trajectories that never enter the system and are reflected instead directly at the tunnel barrier \cite{whitney07,kuipers09}.  For example for the second moment, along with the diagrams in \frefs{secondmoment} and \ref{secondmomenttunnel}, the diagrams in \fref{secondmomentreflection} are also now possible.

Along with the additional possibilities, the tunneling phases now become particularly important \cite{whitney07}.  Looking in detail at the tunnel barrier drawn in \fref{secondmomentreflection}(a), where the incoming channel $i_2$ is identical to the outgoing channel $o_1$, both solid trajectories tunnel through while both dashed, complex conjugated, trajectories are reflected, one on each side of the barrier.  If $\rho$ is the reflection amplitude of the barrier and $\tau$ the transmission amplitude then the four trajectories of the encounter at the barrier give the factor $(\tau\rho^{*})^{2}$ which in the semiclassical limit is equal to $-p(1-p)$ with an additional minus sign arising from the quantum mechanical tunneling \cite{whitney07}.  With equal tunneling probabilities in every channel, the diagrams in \frefs{secondmomentreflection}(a) or (d) then give a total semiclassical contribution of
\begin{equation}
-4p(1-p)\NL\frac{p^2\NL^2}{p^2N^2} = -4p(1-p)N\zeta_1^3 .
\end{equation}

In \frefs{secondmomentreflection}(b) or (e), the tunneling and reflecting trajectories at the tunnel barrier pair a trajectory stretch with a complex conjugated one giving the standard factor of $p(1-p)$ while in \frefs{secondmomentreflection}(c) or (f) none of the trajectories ever enter the system.  Combined they give
\begin{equation}
4p(1-p)N\zeta_1^2+(1-p)^2N\zeta_1 ,
\end{equation}
while the diagrams in \frefs{secondmoment} and \ref{secondmomenttunnel} provide the contribution
\begin{equation}
p(p-2)N\zeta_1^4+2p^2N\zeta_1^3+4p(1-p)N\zeta_1^4 ,
\end{equation}
since now all the channels are in the first lead.  We can then, using $\zeta_1(1-\zeta_1)=\xi$, write the leading order term of the second moment of the reflection eigenvalues as
\begin{equation}
\label{leadingreflectsecond}
\left \langle \Tr\left[\boldsymbol{r}^\dagger \boldsymbol{r}\right]^2 \right \rangle= N\zeta_1 +pN\xi(p-2+2\xi-3p\xi) + O(1) .
\end{equation}
Through the unitarity condition, we have
\begin{equation}
\boldsymbol{r}^{\dagger}\boldsymbol{r} + \boldsymbol{t}^\dagger \boldsymbol{t} = I_{\NL}, \qquad \Tr\left[\boldsymbol{r}^\dagger \boldsymbol{r}\right]^2 = \NL -2\Tr\left[\boldsymbol{t}^\dagger \boldsymbol{t}\right] + \Tr\left[\boldsymbol{t}^\dagger \boldsymbol{t}\right]^2 ,
\end{equation}
which is satisfied by the leading order second moments in \eref{leadingtranssecond} and \eref{leadingreflectsecond} since the leading order first moment of the transmission eigenvalues from \eref{leadingtransfirst} is $pN\xi$.

\subsection{Auxiliary trees}

For the moments of the reflection eigenvalues, we first let $f$ and $\fh$ be generating functions which count all trees including their semiclassical contributions which start inside the system from an incoming or outgoing direction respectively (without being rooted in a channel).  Due to the symmetry, we actually have $f=\fh$ but we will treat them separately for now.  The $f$ trees, which now include diagrams like \frefs{secondmomentreflection}(a) and (b) with the channel $i_1$ removed, initially start from a link connecting to an $l$-encounter which is followed by $l$ subtrees of type $f$ and $l-1$ trees of type $\fh$.  In general, and as illustrated in \fref{secondmomentreflection}, we can now allow any of the links around an encounter node to tunnel straight into the lead and be removed in the graphical representation.  

\begin{figure}
  \centering
  \includegraphics[width=0.82\textwidth]{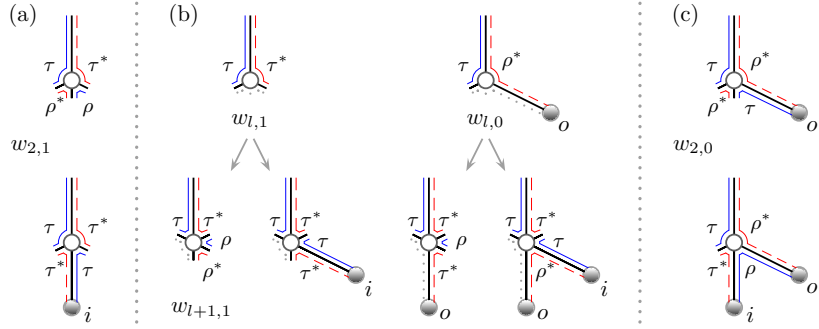}
  \caption{For the auxiliary generating functions $w$ where the first link after the encounter node tunnels directly into the lead, we need to track exactly which trajectories tunnel through or are reflected at the barrier.  For $w_{l,1}$ the last link tunnels so that, depending on whether the middle link tunnels or is reflected, there are two possible configurations in $w_{2,1}$ as shown in (a).  For $w_{l,0}$ the last link is reflected so that again there are two configurations in $w_{2,0}$ shown in (c).  Starting from all the configurations in $w_{l,1}$ and $w_{l,0}$ we can add an additional two links on the right of the encounter.  If the last tunnels into the lead, the two possibilities for the other generate the configurations in $w_{l+1,1}$ as depicted in (b) while if the last link is reflected we obtain $w_{l+1,0}$ and the recursions in \eref{wrecursions}.}
  \label{auxiliarytrees}
\end{figure}

Counting the possibilities recursively would be straightforward, but for the quantum mechanical tunneling phases which actually depend on how many adjacent links are removed together.  To keep track of these phases we introduce auxiliary generating functions $w_{l,\alpha}$ and $\wh_{l,\alpha}$ which count the contributions below the encounter (and not the top link of $f$) and where the encounter is in a particular channel in the lead with tunneling probability $p$.  The subscript $\alpha=1$ represents that the last of the subtrees tunnels straight into the lead and is removed from the diagram and $\alpha=0$ represents that it is reflected.  In $w$ we ensure that the first subtree tunnels into the lead while in $\wh$ we ensure it is reflected.  The smallest encounter is $l=2$ for which the central tree can either tunnel or be reflected so, as in \frefs{auxiliarytrees}(a) and (c), we have
\begin{equation}
\fl w_{2,1} = \tau \rho^{*} \rho \tau^{*} r^3 + (\tau \tau^{*})^{2} r^2 \fh , \qquad w_{2,0} = (\tau \rho^{*})^{2} r^2 f + \tau \tau^{*} \rho \rho^{*} r f\fh ,   
\end{equation}
and
\begin{equation}
\fl \wh_{2,1} = (\rho \tau^{*})^{2} r^2 f + \rho \rho^{*} \tau \tau^{*} rf\fh , \qquad \wh_{2,0} = \rho \tau^{*} \tau \rho^{*} r f^2 + (\rho \rho^{*})^2 f^2\fh ,    
\end{equation}
where for simplicity we have included an encounter entirely reflected at the lead in $\wh_{2,0}$ which needs to be remembered later.  With the symmetry we also have $w_{l,0}^{*} = \wh_{l,1}$ and we can simplify by setting $\tau \tau^{*} = p$, $\rho \rho^{*} = (1-p)$ and $(\tau\rho^{*})^{2}=(\rho\tau^{*})^{2}=-p(1-p)$.

Since we know the behaviour of the subtrees at the edges, we can recursively generate $w_{l+1,a}$ from $w_{l,a}$ by adding two more subtrees on the right and allowing both possibilities for the first.  Tracking the trajectories as in \frefs{auxiliarytrees}(b) we obtain
\begin{eqnarray}
\label{wrecursions}
w_{l+1,1} &=& w_{l,1}\left[\rho \rho^{*} r^2 + \tau \tau^{*} r \fh \right] +  w_{l,0}\left[\frac{\tau^{*} \rho \tau^{*}}{\rho^{*}} r^2 + \tau \tau^{*} r\fh \right] , \nonumber \\
w_{l+1,0} &=& w_{l,0}\left[\tau \tau^{*} r f + \rho \rho^{*} f\fh \right] +  w_{l,1}\left[\frac{\rho^{*} \tau \rho^{*}}{\tau^{*}} rf + \rho \rho^{*} f\fh \right] ,
\end{eqnarray}
where the terms with fractions involve replacing a reflecting trajectory by a transmitting one or vice versa, and have the values $\tau^{*} \rho \tau^{*}/\rho^{*} = -p$ and $\rho^{*} \tau \rho^{*}/\tau^{*} = (p-1)$.  Because this recursion does not depend on whether the first subtree tunnels or is reflected, we have the same equations for $\wh$.  We wish to sum over encounters of all sizes, so if we set $w_{\alpha}=\sum_{l=2}^{\infty}w_{l,\alpha}$, and likewise for $\wh$ we obtain the following coupled equations
\begin{eqnarray} \label{wsimprecursions}
\fl w_{1} - p(1-p)r^3 - p^2r^2\fh &=& w_{1}\left[(1-p)r^2 + pr\fh\right] + w_{0}pr\left[\fh - r\right] ,\nonumber \\
\fl w_{0} - p(1-p)rf(\fh-r)&=& w_{0}\left[prf + (1-p)f\fh \right] + w_{1}(1-p)f\left[\fh-r\right].
\end{eqnarray}
We obtain a similar equation for $\wh$ and the solutions, with $f=\fh$, are
\begin{eqnarray}
w_{1} &=& \frac{pr^2[r(1-p-f^2)-pf]}{1-(1-p)r^2-2prf-(1-p-r^2)f^2} , \nonumber \\
w_{0} &=& \wh_{1} = \frac{prf(p-1)(r-f)}{1-(1-p)r^2-2prf-(1-p-r^2)f^2} , \nonumber \\
\wh_{0} &=& \frac{f^2(1-p)[f(1-p-r^2)+pr]}{1-(1-p)r^2-2prf-(1-p-r^2)f^2} .\nonumber \\
\end{eqnarray}

\subsection{Tree recursions}

To obtain the generating function $f$ we simply add a link to these contributions and sum over all channels in the lead
\begin{equation}
\label{frefleqn}
\fl \frac{f}{y} =  r\sum_{i=1}^{\NL} p_i - \sum_{l=2}^\infty \sum_{i=1}^{N} f^{2l-1} + \sum_{l=2}^\infty \sum_{i=\NL+1}^{N}(1-p_i)^{l}f^{2l-1} + \sum_{i=1}^{\NL} (w_1 + 2w_0 + \wh_0) ,
\end{equation}
where the sum over the $(1-p_i)^l$ terms in the first lead are already included in $\wh_0$.  With equal tunneling probabilities in all the channels, this is
\begin{equation}
\label{frefleqnsimp}
\frac{f}{1-f^2} =  \frac{f(1-p)[1-\zeta_1f^2(1-p)]}{1-f^2(1-p)} + \zeta_1(rp+w_1 + 2w_0 + \wh_0) ,
\end{equation}
which leads to a quartic for $f$ whose expansion is
\begin{eqnarray}
\label{fexpansion}
\fl f &=&  \zeta_1 r + \xi\left(1-p-\zeta_1(1-2p)\right)r^3 +  \\
\fl  & & {} + \xi\left((1-p)^2-\zeta_1(1-p)(1-2p)-\xi(1-p)(2-5p)+2\zeta_1\xi[1-5p(1-p)] \right)r^5 \ldots \nonumber
\end{eqnarray}
The second term now corresponds to the sum of the diagrams in \frefs{secondmoment}(d) and (f) and \frefs{secondmomenttunnel}(d) and (f) along with two new possibilities like the diagram in \fref{secondmomentreflection}(d) and one diagram like \fref{secondmomentreflection}(e).

\subsection{Leading order moments}

To move from the generating function $f$ to the leading order moment generating function for the reflection eigenvalues $R_0$ we need to add a channel to the top of the diagrams in $f$ and also allow the top link in $f$ to tunnel straight into the lead.  Since we know whether both the outside subtrees of $w$ and $\wh$ tunnel or are reflected we can simply replace their top lead by a tunneling one and replace the corresponding transmitting and reflecting eigenvalues of the trajectories
\begin{equation}
\label{R0eqn1}
\fl NR_0 = \sum_{i=1}^{\NL}rp_i f + \sum_{i=1}^{\NL}r\left(\frac{\rho\rho^{*}}{\tau\tau^{*}}w_{1}+\frac{\rho\tau^{*}}{\tau\rho^{*}}w_{0}+\frac{\rho\tau^{*}}{\tau\rho^{*}}\wh_{1}+\frac{\tau\tau^{*}}{\rho\rho^{*}}\wh_{0}\right) + \sum_{i=1}^{\NL}r^2(1-p_i) ,
\end{equation}
where the last term is the contribution to the first moment from a trajectory pair that never enters the system.  We have $\rho\tau^{*}/(\tau\rho^{*}) = -1$ so that in total we obtain
\begin{equation}
\label{R0eqngeneral}
NR_0 = \sum_{i=1}^{\NL}\frac{r\zeta_1[r(1-p_i-f^2)+p_{i}f]}{1-(1-p_i)r^2-2p_irf-(1-p_i-r^2)f^2} .
\end{equation}

With equal tunneling probabilities, we again obtain a quartic for $R_0$ which can be simplified to
\begin{eqnarray}
\label{R0shiftquartic}
\fl \left[4(s-1)(1-p)+sp^2\right]\left(\tilde{R}_0^4 + \tilde{R}_0^3 + s\xi\tilde{R}_0^2\right) 
+ sp^2\tilde{R}_0^3 \nonumber \\
\fl + \left[(s+p-1)(1-s+sp)+sp^2\xi\right]\tilde{R}_0^2 + s(1+s)p^2\xi \tilde{R}_0+s^2p^2\xi^2 = 0 ,
\end{eqnarray}
where $\tilde{R}_0= (1-s)R_0 - \zeta_1s$ removes the $\zeta$ dependence.  The expansion provides
\begin{eqnarray}
\fl R_0 & = & \zeta_1(s+s^2+s^3) -p\xi s + p\xi\left(p-2+2\xi-3p\xi\right)s^2 \nonumber \\
\fl & & {} - p\xi\left(p^2-3p+3-(6-15p+8p^2)\xi+(6-21p+17p^2)\xi^2\right)s^3 +\ldots 
\end{eqnarray}
The coefficient of $s^2$ is the same as \eref{leadingreflectsecond} and unitarity can be checked against \eref{T0expansion}.

With equal leads $\zeta_1=1/2$, we end up with the same quadratic $rf^2-2f+r=0$ that we had for the transmission, while the moment generating function is
\begin{equation}
\label{R0quadraticmgf}
R_0=\frac{s(2-p)\left((2-p)(1-s)+p\sqrt{1-s}\right)}{2(s-1)\left(4(s-1)(1-p)+sp^2\right)} .
\end{equation}
Transforming to the probability distribution of the reflection eigenvalues, we have the same density as in \eref{semiclassicalprobdensity} but with $Z$ replaced by $1-Z$ which is the same mapping as between the transmission and reflection eigenvalues.

\subsection{One perfect lead} \label{perfectreflection}

Since the start and end channels are all in the same lead, is it straightforward to make the other lead transparent.  Keeping $p_i=p$ in the first lead and setting $p_i=1$ in the second means that we remove the sum over the channels in the second lead in \eref{frefleqn} and set $1/y= N(p\zeta_1+\zeta_2)$.  This changes \eref{frefleqnsimp} to
\begin{equation}
\label{frefllead2perfect}
\frac{f}{1-f^2} =  \zeta_1 f(1-p) + \zeta_1(rp+w_1 + 2w_0 + \wh_0) ,
\end{equation}
while \eref{R0eqngeneral} remains the same.  This leads to a cubic for $f$ and $R_0$, where again using $\tilde{R}_0= (1-s)R_0 - \zeta_1s$ leads to the simpler form
\begin{eqnarray}
\label{R0shiftcubic}
(s+p-1)\tilde{R}_0^3 + sp(1+\zeta_2)\tilde{R}_0^2 \nonumber \\
+ s\left[(1-s+sp)\zeta_2+(s+p-1)\xi\right]\tilde{R}_0 + s^2p\zeta_2\xi = 0 .
\end{eqnarray}

Of course we could instead allow the first lead to be transparent and set $p_i=1$ there and $p_i=p$ in the second lead.  In this case the encounters can no longer partially enter the lead so the number of possible diagrams is drastically reduced to just those where the encounter enters the lead fully.  We actually then have almost exactly the same recursions as when there are no tunnel barriers in either lead, but just with the minor corrections to the survival probabilities of the encounters and links.  As in \cite{bk11} we obtain
\begin{equation}
\label{frefllead1perfect}
\frac{f}{Ny} =  r \zeta_1 - \sum_{l=2}^{\infty}f^{2l-1} + r\zeta_1\sum_{l=2}^{\infty}r^{l-1}f^{l-1} + \zeta_2\sum_{l=2}^{\infty}(1-p)^{l}f^{2l-1} ,
\end{equation}
where the last term is due to the change in the survival probability of the encounters while the links provide $1/(Ny) = \zeta_1 + p \zeta_2$.  Simplifying we get
\begin{equation}
\label{frefllead1perfectsimp}
\frac{f}{1-f^2} =  \frac{r \zeta_1}{1-rf} + \frac{(1-\zeta_1)(1-p)f}{1-(1-p)f^2} ,
\end{equation}
where the last term is the correction due to the tunnel barrier.  The moment generating function is still given by
\begin{equation}
\label{R0lead1perfect}
R_0= \frac{r \zeta_1 f}{1-rf} ,
\end{equation}
which leads directly to the cubic
\begin{eqnarray}
\label{R0cubic}
(s-1)(s+p-1)R_0^3 + s\left[\zeta_1(3s+2p-3)-p\right]R_0^2 \nonumber \\
+ s\zeta_1\left[\zeta_1(3s+p-1)-p\right]R_0 + s^2\zeta_1^3 = 0 .
\end{eqnarray}
Shifting the generating function as before, $\tilde{R}_0= (1-s)R_0 - \zeta_1s$, we then obtain exactly \eref{R0shiftcubic} but with $\zeta_2$ replaced by $\zeta_1$, which leaves $\xi$ unchanged.  Swapping $\zeta_1$ and $\zeta_2$ just means we are considering the moments of the reflection eigenvalues of the second lead 
\begin{equation}
R'(s) = \sum_{n=1}^{\infty}s^n \Tr\left[{\boldsymbol{r}'}^\dagger \boldsymbol{r}'\right]^n = NR'_0 + R'_1 + \ldots
\end{equation}
while the unitarity condition $\boldsymbol{r}'{\boldsymbol{r}'}^\dagger + \boldsymbol{t} \boldsymbol{t}^\dagger = I_{\NR}$ ensures that
\begin{equation}
R(s)-\frac{\NL s}{1-s} = R'(s)-\frac{\NR s}{1-s} ,
\end{equation}
so this much simpler treatment provides the same generating function \eref{R0shiftcubic} as the full auxiliary tree combinatorics when one lead is transparent.  This leading order generating function should also arise as the first term of an asymptotic expansion of the recently derived RMT probability distribution \cite{vk12} when the remaining tunneling probabilities are set equal.

\subsection{Summary of different results}

A summary of the different restrictions considered above and the resulting moment generating functions is given in Table~\ref{reflgenfuncts}.

\Table{\label{reflgenfuncts}Leading order generating functions for the moments of the reflection eigenvalues for different restrictions on the tunneling probability and number of channels in each lead.}
\begin{tabular}{cccc}
\br
Tunneling probability & Tunneling probability & Equal number & Equation \\
in the first lead & in second lead & of channels & for $R_0$ \\
\mr
$p$ & $p$ & no & \eref{R0shiftquartic} \\
$p$ & $p$ & yes & \eref{R0quadraticmgf} \\
$p$ & 1 & no & \eref{R0shiftcubic} \\
1 & $p$ & no & \eref{R0cubic} \\
\br
\end{tabular}
\endTable

\section{Moments of the Wigner delay times} \label{wigner}

The tree recursions for the moments of the reflection eigenvalues can easily be modified to obtain energy dependent generating functions like 
\begin{equation} \label{Cepsdefeqn}
C(\epsilon,n) = \frac{1}{N}\Tr \left[S^{\dagger}\left(E-\frac{\epsilon\mu\hbar}{2}\right)+S\left(E+\frac{\epsilon\mu\hbar}{2}\right)\right]^{n} ,
\end{equation}
which are related to other physical observables like the density of states of Andreev billiards and the moments of the Wigner delay times.  The generating function $G(s) = \sum_{n=1}^{\infty} s^n C(\epsilon,n)$ can be obtained by simply considering the reflection eigenvalues with a single lead $\zeta_1=1$ and including the energy difference.  This changes the encounter and link contributions to
\begin{equation}
-N(1-(1-p)^l - la), \qquad y^{-1} =  N(p - a) ,
\end{equation}
respectively with $a=\rmi \epsilon$.  The tree generating function becomes  
\begin{equation}
\label{fandreeveqnsimp}
\frac{f}{1-f^2} =  f(1-p) +\frac{af}{[1-f^2]^{2}} + rp+w_1 + 2w_0 + \wh_0 ,
\end{equation}
while $w$ and $\wh$ remain unchanged.  The terms can be combined and simplified to
\begin{equation}
\label{fandreeveqnsimpler}
\frac{f(1-f^2-a)}{[1-f^2]^2} = \frac{pr+(1-p-r^2)f}{1-(1-p)r^2-2prf-(1-p-r^2)f^2} .
\end{equation}
The leading order generating function $G_0$ is still given by \eref{R0eqngeneral} with $\zeta_1=1$ so that we obtain a quartic equation for $f$ and $G_0$.

We show how to use the energy dependence to obtain the leading order contribution to the density of states of Andreev billiards next in \sref{andreev} and concentrate here on the moments of the Wigner delay times.  The delay times are the eigenvalues of the Wigner-Smith matrix \cite{wigner55,smith60}
\begin{equation}
Q=\frac{\hbar}{\rmi}S^{\dagger}(E)\frac{\rmd S(E)}{\rmd E} ,
\end{equation}
and are a measure of the time spent inside the scattering cavity.  Their moments can be obtained \cite{bk10,bk11,lehmannetal95} through the correlation functions  
\begin{equation}
D(\epsilon,n)= \frac{1}{N}\Tr \left[S^{\dagger}
\left(E-\frac{\epsilon\mu\hbar}{2}\right)S\left(E+\frac{\epsilon\mu\hbar}{2}\right)
-I\right]^n ,
\label{Depseqn}
\end{equation}
by differentiating
\begin{equation}
\label{motDepseqn}
\Tr \left[Q\right]^{n} = \frac{1}{(\rmi\mu)^{n}n!}\frac{\rmd^{n}}{\rmd \epsilon^{n}} D(\epsilon,n)\Big\vert_{\epsilon=0} .
\end{equation}
If we denote the moment generating function by
\begin{equation}
M(s)= \sum_{n=1}^{\infty} \mu^n s^n \left \langle \Tr \left[Q\right]^{n} \right \rangle ,
\end{equation}
then we can obtain an expansion for the leading order lower moments by expanding \eref{Depseqn} binomially and simply substituting the correlation functions $C(\epsilon,n)$
\begin{equation} \label{M0expansion}
\fl M_0(s)= s + \frac{2}{p}s^2 + \frac{6}{p^2}s^3 + \frac{2(p+10)}{p^3}s^4 + \frac{10(2p+7)}{p^4}s^5 + \ldots
\end{equation}

\subsection{Tree recursions}

However to obtain the full moment generating function we need to account for the identity matrix in the brackets in \eref{Depseqn} that creates the difference from \eref{Cepsdefeqn}.  As was the case without tunnel barriers \cite{bk10,bk11}, this matrix can be thought of as coming from diagonal trajectory pairs which just travel directly from the incoming to outgoing channels with no energy difference.  With tunnel barriers, such diagonal pairs can be formed whenever an encounter goes into the lead and an $\fh$ subtree remains, as long as the other trajectories either side end in the lead so that we have stubs either side.  The subtree contribution already includes a diagonal pair and to subtract the identity matrix we just subtract the contribution of a diagonal pair with no energy difference ($a=0$).  For every $\fh$ subtree surrounded by two stubs we then replace its contribution by $\fh-r/p$.  This breaks the symmetry between $f$ and $\fh$ and we look at their two generating functions separately.

The auxiliary tree recursions for $f$, make it easy to see when a $\fh$ subtree is added with a stub either side.  Modifying the recursions appropriately we find
\begin{eqnarray} \label{wsimpnewrecursions}
\fl w_{1} - p^2r^2(\fh-r) &=& pr(\fh-r)(w_{1} + w_{0}) , \nonumber \\
\fl w_{0} - p(1-p)rf(\fh-r)&=& w_{0}\left[pr\fh + (1-p)f\fh \right] + w_{1}(1-p)f(\fh - r) ,
\end{eqnarray}
and similar equations for $\wh$.  This leads to the equation for $f$
\begin{equation}
\label{fwignereqnsimpler}
\fl\frac{f(1-f\fh-a)}{[1-f\fh]^2} = \frac{pr+(1-p-pr^2)f}{1+pr^2-pr(f+\fh)-(1-p-pr^2)f\fh-pr^3f} .
\end{equation}

For $\fh$ however, because we add two new subtrees on the right in the auxiliary tree recursions, it is not so straightforward to see when a $\fh$ subtree is surrounded by two stubs.  Instead we can simply add either a $\fh$ subtree or a stub to either side of the auxiliary trees for $f$.  Doing so we find
\begin{equation}
\label{fhwignereqnsimpler}
\fl\frac{\fh(1-f\fh-a)}{[1-f\fh]^2} = \frac{pr(1+r^2)+(1-p-pr^2)\fh}{1+pr^2-pr(f+\fh)-(1-p-pr^2)f\fh-pr^3f} ,
\end{equation}
which, by multiplying \eref{fwignereqnsimpler} by $\fh$ and \eref{fhwignereqnsimpler} by $f$, leads to the simple relation $\fh=(1+r^2)f$ and a quartic equation for $f$ or $\fh$.

\subsection{Leading order moments}

To obtain the generating function
\begin{equation}
L(s) = \sum_{n=1}^{\infty}s^n D(\epsilon,n) 
\end{equation}
we then either root $f$ in an incoming channel or tunnel straight into the auxiliary trees to obtain
\begin{equation}
\label{L0eqn1}
\frac{L_0}{r} = pf - w_{1} - w_{0} - \wh_{1} + \frac{p}{(1-p)}\wh_{0} - pr ,
\end{equation}
where we included the correction $-rw_1/p$ from the diagonal pair above the auxiliary $w_1$ trees which have stubs on either side.  The last term stems from a pair of trajectories that never enter the system.  Here the $w$ terms correspond to the auxiliary trees from the recursions for $f$ as in \eref{wsimpnewrecursions} so that finally we obtain
\begin{equation}
\label{L0eqn2}
L_0 = \frac{pr(f-r-2r^2f+rf\fh)}{1+pr^2-pr(f+\fh)-(1-p-pr^2)f\fh-pr^3f} ,
\end{equation}
and the quartic
\begin{eqnarray}
\label{Lquartic}
 p^2(1+s)L_0^4 + 2aps(1+s)(2-p)L_0^3 \nonumber \\
 {} + s\left[(ap-a+p)(a-p) + ap(ap-2p+4)s + a^2p^2s^2\right]L_0^2 \nonumber \\
 {}  + a^2p^2s^2(1+2s)L_0 + a^2p^2s^3 = 0 .
\end{eqnarray}

For the $n$th moment of the delay times we then differentiate $n$ times (and divide by $n!$) which can be achieved by simply transforming $s\to s/a$.  Setting the energy difference $a$ to 0 then provides the generating function 
\begin{equation}
\label{Meqn}
p{M_0}^3+(4s-2ps-p){M_0}^2+ps^2{M_0}+ps^2=0 .
\end{equation}
Expanding $M_0$ gives the terms in \eref{M0expansion} directly when we pick the solution whose low moments agree with the semiclassical diagrams.  The probability distribution of the delay times has previously been obtained from RMT including the leading order distribution in the limit of a large number of channels in the lead \cite{sss01}, which is also derived from a cubic equation.  Our result in \eref{Meqn} should then be the Hilbert transform of the result in \cite{sss01}.

\section{Andreev billiards} \label{andreev}

Andreev billiards are systems where the (single) normal conducting lead has been replaced by a superconductor to make a closed ballistic system trapping electrons which are converted to holes (and vice versa) when they interact with the superconductor.  Despite being closed systems, they can actually be treated semiclassically with similar methods to transport quantities. Interference effects, as exemplified by encounters between trajectory sets in the semiclassical tree recursions, lead to an energy region where the system supports no quantum states and a hard gap in the density of states \cite{kuipersetal10,melsenetal96}.  The semiclassical treatment of Andreev billiards was detailed in \cite{kuipersetal11} so we just highlight the incorporation of tunnel barriers here.

In the scattering approach \cite{beenakker05}, the density of states \cite{ihraetal01}
\begin{equation}
d(\epsilon) = 1 + 2\Imag \sum_{n=1}^{\infty}\frac{(-1)^n}{n} \frac{\partial C(\epsilon,n)}{\partial \epsilon} 
\end{equation}
involves a slowly converging series of the correlation functions $C(\epsilon,n)$ defined in \eref{Cepsdefeqn} and generated in \sref{wigner}.  For the density of states however we would rather directly obtain (again with $a=\rmi \epsilon$)
\begin{equation}
H(s) = \sum_{n=1}^{\infty} \frac{s^n}{n} \frac{\partial C(\epsilon,n)}{\partial a} ,
\end{equation}
where to divide by $n$ we can generate the trees without rooting them in a specific channel, as is automatically the case when looking beyond leading order as in \ref{subleading}.  At leading order \cite{bk11}, we can join $2l$ subtrees at a single point to make an encounter and divide by the rotational symmetry factor of $2l$,
\begin{eqnarray}
\fl \tilde{K}_0 &=& -\sum_{l=2}^{\infty} \frac{(1-la)f^{2l}}{2l}+f\sum_{l=2}^{\infty} \frac{w_{l,1} + 2w_{l,0} + \wh_{l,0}}{2l} \nonumber \\
\fl & & {} +r\sum_{l=2}^{\infty} \frac{\left((1-p)^2w_{l,1}-2p(1-p)w_{l,0}+p^2\wh_{l,0}/(1-p)\right)}{2p(1-p)l} ,
\end{eqnarray}
to overcount all unrooted trees by the number of their encounter nodes.  The last two terms correspond to allowing the encounter to touch the lead with the top link either being reflected or tunneling straight through, while the $(1-p)^l$ term from the encounter is still included in $\wh_{l,0}$.  To perform the sums divided by $l$ we can include powers of $q$ in the recursion relations for $w$ and $\wh$ in \eref{wrecursions} and find and solve the corresponding coupled equations for $\sum_{l=2}^{\infty}q^{l-1}w_{l,1}$and $\sum_{l=2}^{\infty}q^{l-1}w_{l,0}$.  Integrating with respect to $q$ and setting it equal to 1 then provides the required sums and we obtain
\begin{eqnarray}
\fl 2\tilde{K}_0 &=& \ln(1-f^2)+\frac{af^4}{1-f^2} - \ln\left[1+(p-1)(r^2+f^2)+rf(rf-2p)\right] \nonumber \\ 
\fl & & {} +(p-1)r^2+fp(f-2r) .
\end{eqnarray}

We can also fuse two subtrees (with at least one encounter node each) together, 
\begin{equation}
2\tilde{K}'_0 = (p-a)\left(f-\frac{rp}{p-a}\right)^2 - \frac{p^2r^2}{(p-a)} - (1-p)r^2 ,
\end{equation}
to overcount all unrooted trees by the number of their internal edges, which is one less than the number of encounter nodes.  The last two terms are corrections for the first moment.  The difference $\tilde{K}_0-\tilde{K}'_0$ then counts the unrooted trees exactly once and we obtain the desired generating function
\begin{equation}
\fl 2K_0 = 2\left(\tilde{K}_0 - \tilde{K}'_0\right) = \ln\left(\frac{1-f^2}{1+(p-1)(r^2+f^2)+rf(rf-2p)}\right)+\frac{af^2}{1-f^2} ,
\end{equation}
which provides the leading order terms of $C(\epsilon,n)/n$.  Differentiating implicitly with respect to the energy difference, or $a$, gives the quartic satisfied by $H_0$
\begin{eqnarray}
\label{Hquartic}
\fl ap^2(s-1)^2\left(aH_0^4 - 2H_0^3\right) + a^2p(1-s)(2+p+2s-3sp)H_0^3 \nonumber \\
\fl {} + \left[(1-s)(p+a) + pas\right]^2\left(H_0^2+H_0\right) + 2pa(s-1)(2+p-ps-a-as+pas)H_0^2 \nonumber \\
\fl {}  + 4pa(s-1)H_0 -sp^2 = 0 .
\end{eqnarray}
\begin{figure}
  \centering
  \includegraphics[width=0.6\textwidth]{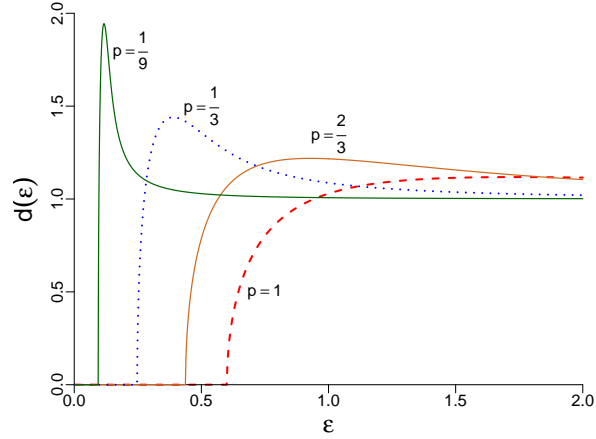}
  \caption{The leading order contribution to the density of states of Andreev billiards with a tunnel barrier between the superconductor and the cavity for different tunneling probabilities $p$.}
  \label{andreevdos}
\end{figure}

We can now make the substitution
\begin{equation}
H_0(s=-1) = \frac{\rmi W_0(\epsilon) -1}{2} ,
\end{equation}
and replace $a=\rmi \epsilon$ to obtain the leading order contribution to the density of states $d_0(\epsilon)=-\Imag W_0(\epsilon)$ through the quartic
\begin{eqnarray}
\label{Wquartic}
p^2\epsilon^2W_0^4 + 4p^2\epsilon W_0^3 +\left(4p^2+p^2\epsilon^2+4p\epsilon^2-4\epsilon^2\right)W_0^2 \nonumber \\
{} + 4p^2\epsilon W_0 + 4p\epsilon^2-4\epsilon^2 = 0 .
\end{eqnarray}
This is identical to the RMT result \cite{melsenetal96} and some example density of states for various transparencies $p$ of the tunnel barrier are depicted in \fref{andreevdos}.

\section{Conclusions} \label{concs}

Adding tunnel barriers to the leads of a chaotic cavity, to better approximate realistic experimental settings, leads to a wealth of additional correlated trajectory structures that contribute in the semiclassical approach.  These centre around encounters which partially enter the lead and for transmission through the cavity can be incorporated into the previous treatment with perfect leads \cite{bhn08,bk11} relatively easily.  However the increase in complexity meant we could only obtain closed form generating functions for the moments of the transmission eigenvalues at leading order if we restricted our attention to the case where the tunneling probabilities are equal for each channel in each lead as in \eref{Phiquarticdiffleads}.  Neatly, this is a complementary regime to the leading order RMT results which allow different tunneling probabilities in each channel as long as the leads are identical \cite{bb96}.

At subleading order, the previous treatment \cite{bk11} could likewise be easily extended as in \ref{subleading}, but apart from a computational expansion for the low moments, a generating function for all moments could only be obtained in the even more restricted case of equal leads and equal tunneling probability for all channels as in \eref{transsubleading}.  This suggests the same case would need to be considered at higher orders in inverse channel number, limiting the general semiclassical approach to low orders or low moments.

For reflection quantities, even more diagrammatic possibilities occur but of greater importance are the new quantum mechanical phases that occur at the tunnel barrier \cite{whitney07}.  These phases are essential for the unitarity of the semiclassical approach and we introduced new combinatorial structures to be able to account for them.  With these structures, the leading order moments of the reflection eigenvalues in \eref{R0shiftquartic} could be obtained.  A connection to RMT arises when making one lead perfect while leaving a tunnel barrier in the other lead as in \sref{perfectreflection} since the probability distribution of the reflection eigenvalues for such a cavity has recently been derived \cite{vk12}.  Integrating this distribution and analysing the asymptotics of a large number of channels should lead back to our moment generating function \eref{R0shiftcubic}.  However in the case of one perfect and one tunneling lead, from the unitarity of the scattering matrix, we can actually obtain this moment generating function directly from the treatment with two perfect leads.  This may suggest that the RMT probability distribution for two tunneling leads could be much more complex.  

Including energy dependence in the new auxiliary structures allowed us to obtain the density of states of Andreev billiards as well as the moment generating function for the Wigner delay times.  The full probability density of the delay times in known from RMT \cite{sss01} and it would be interesting to see if asymptotic analyses of its integral, like those performed without tunnel barriers \cite{ms11,ms12}, would recreate the generating function in \eref{Meqn}.  Also treated in the asymptotic expansions of \cite{ms11,ms12} were the transport moments of quantum dots connected to superconductors and their corresponding symmetry classes \cite{az97}.  Semiclassically the conductance through such systems \cite{wj09,ekr11} involves different tree species and the auxiliary structures we introduced for the moments of the reflection amplitudes could be useful for investigating the higher moments of these superconducting symmetry classes.

The approximations used for the semiclassical diagrammatic rules are valid in the regime where the time trajectories spend in the cavity is shorter than the Heisenberg time.  Without tunnel barriers, this translates to $N>1$ which, seeing as there must be at least one channel in each lead to allow transport, is automatically the case (the approximations actually seem to hold all the way to $N=1$ for quantities with a single lead).  Effects due to correlations above the Heisenberg time can then safely be neglected.  However, with tunnel barriers the condition becomes $pN>1$ so that even with a large number of channels, a weak enough tunneling probability $p$ can allow Heisenberg time effects to appear.  How to treat such long time trajectory correlations semiclassically is currently unknown, although they have been treated indirectly in the related problem of the spectral statistics of closed chaotic systems using pseudo-orbit correlations and resummation imposing unitarity \cite{heusleretal07,mulleretal09}. 

For the diagrammatic rules, the dwell time that trajectories typically spend in the cavity should also be much larger than the Ehrenfest time.  As the Ehrenfest time becomes comparable, additional diagrammatic structures arise and the semiclassical evaluations become more complicated.  This has limited the treatment without tunnel barriers to low moments and low orders \cite{jw06,br06,wj06,br06b,wk10} although the leading order contribution for all moments has also been obtained \cite{wkr11}.  With tunnel barriers this regime has only been investigated for fewer quantities at lower moments and order \cite{whitney07,petitjeanetal09}, with particularly interesting behaviour arising from the interplay of the tunnel barriers and Ehrenfest time effects for diagrams contributing to the universal conductance fluctuations \cite{waltneretal12}.

\ack{The authors would like to thank Gregory Berkolaiko for helpful discussions and gratefully acknowledge the DFG for funding through FOR 760.}

\appendix

\section{Subleading order for transmission} \label{subleading}

\begin{figure}
  \centering
  \includegraphics[width=0.84\textwidth]{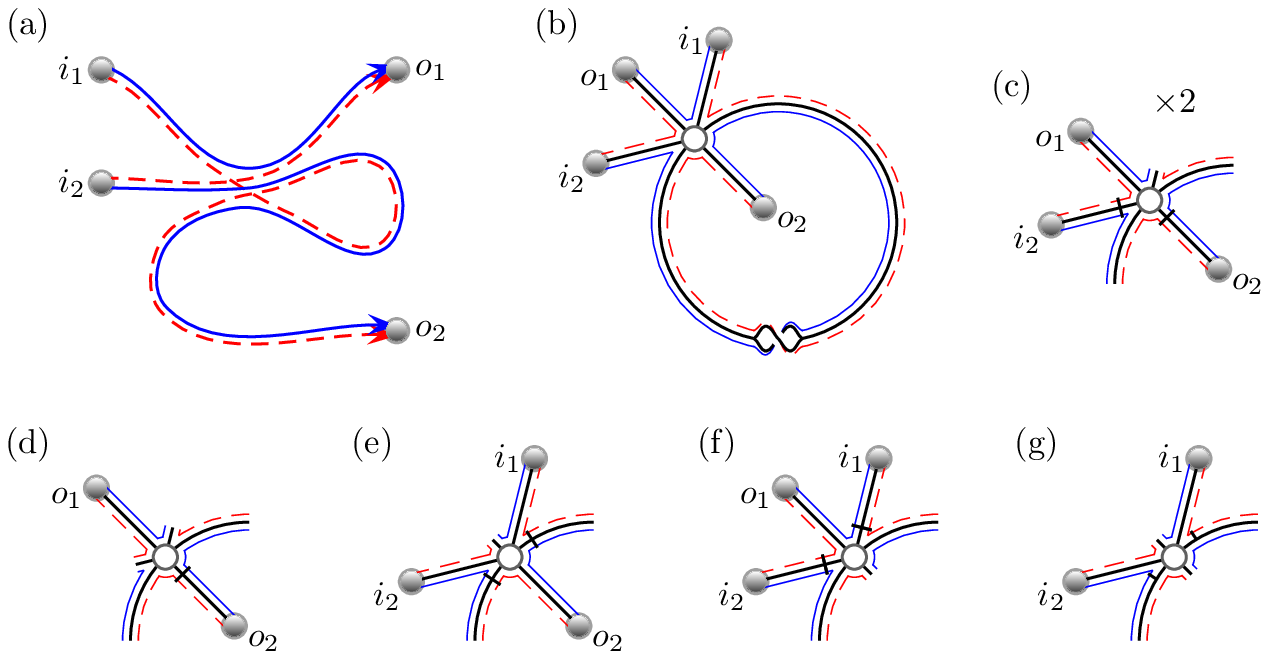}
  \caption{At subleading order, the trajectory quadruplet contributing to the second moment in (a) involves a loop traversed in opposite directions by a trajectory and its partner and so requires time reversal symmetry.  The trajectories can be redrawn around the graph in (b) which is a M\"obius strip with an encounter node where trees connecting to the channels meet.  With tunnel barriers we can allow this encounter node to touch the incoming lead and have one (c) or both (d) of the trajectory stretches going to the incoming channels tunnel directly into the lead.  The rest of the encounter must be reflected at the tunnel barrier so that the remaining trajectory stretches can travel to the other lead.  The encounter may also move into the outgoing lead as in (e)--(g) where the reflected trajectory stretches may now be along the M\"obius loop itself and, as in (g), can become separated in the graphical representation.}
  \label{subleadingtunnel}
\end{figure}

Diagrams at subleading order, like the example in \fref{subleadingtunnel}(a), all involve a loop traversed in opposite directions and therefore require time reversal symmetry.  This loop can be represented graphically as a M\"obius strip, while the rest of the diagram remains as trees which meet the M\"obius strip at encounter nodes \cite{bk11}.  The trajectory quadruplet in \fref{subleadingtunnel}(a) then transform to the graph in \fref{subleadingtunnel}(b).  Each $l$-encounter leads to $l-1$ of each of the leading order trees $\phi$ and $\ph$.  These are arranged alternately around the encounter with an arbitrary number on each side of the M\"obius strip.  If the numbers on each side are odd or even, we correspondingly describe the node as odd or even.  One complications is that to close the M\"obius loop we require an odd number of odd nodes around it.

With tunnel barriers, we can also move the encounters into the incoming or outgoing leads and allow some links from the encounter nodes to tunnel straight into the leads while the remaining parts of the diagram are reflected back into the cavity.  From the diagram in \fref{subleadingtunnel}(a) or (b), we obtain the additional possibilities in \frefs{subleadingtunnel}(c)--(g).  For the semiclassical contribution $B$ of each odd node, for which there are $l-1$ ways of having an odd number of trees on each side, we obtain 
\begin{eqnarray}
\fl \frac{B}{q} &=&  - \sum_{l=2}^\infty \sum_{i=1}^{N} \left(1-(1-p_i)^l\right)(l-1)\phi^{l-1}\ph^{l-1} \nonumber \\
\fl & & {} +\sum_{l=2}\sum_{i=\NL+1}^{N}(l-1)\ph^{l-1}\sum_{k=1}^{l-1} \left(\begin{array}{c}l-1 \\ k \end{array} \right)\phi^{l-1-k}(1-p_i)^{l-k}p_i^kr^k \nonumber \\
\fl & & {} +\sum_{l=2}\sum_{i=1}^{\NL}(l-1)\phi^{l-1}\sum_{k=1}^{l-1} \left(\begin{array}{c}l-1 \\ k \end{array} \right)\ph^{l-1-k}(1-p_i)^{l-k}p_i^kr^k ,
\end{eqnarray}
where we include a factor of $q$ to later count the number of odd nodes and do not include the contribution from the links which form the M\"obius strip itself.  Again the $(1-p)^l$ term corresponds to the $k=0$ terms of both sums, leading to
\begin{eqnarray}
\label{Beqn}
\frac{B}{q} &=& - \frac{N\phi\ph}{(1-\phi\ph)^2} + \sum_{i=\NL+1}^{N}\frac{\ph(1-p_i)[rp_i+\phi(1-p_i)]}{[1-rp_i\ph-\phi\ph(1-p_i)]^2} \nonumber \\
 & & {} +\sum_{i=1}^{\NL}\frac{\phi(1-p_i)[rp_i+\ph(1-p_i)]}{[1-rp_i\phi-\phi\ph(1-p_i)]^2} ,
\end{eqnarray}
There are $l$ ways of arranging an even number of trees on each side, giving a further contribution of 
\begin{eqnarray}
\label{Aeqn}
A &=& \frac{B}{q} - \frac{N\phi\ph}{1-\phi\ph} + \sum_{i=\NL+1}^{N}\frac{\ph(1-p_i)[rp_i+\phi(1-p_i)]}{1-rp_i\ph-\phi\ph(1-p_i)} \nonumber \\
 & & {} +\sum_{i=1}^{\NL}\frac{\phi(1-p_i)[rp_i+\ph(1-p_i)]}{1-rp_i\phi-\phi\ph(1-p_i)} .
\end{eqnarray}

Around the M\"obius strip we arrange an arbitrary number of nodes each separated by a link, account for the rotational symmetry by dividing by the number of nodes, and then ensure we have an odd number of odd nodes \cite{bk11} to obtain the generating function 
\begin{equation}
\fl \tilde{K}_1 = -\frac{1}{2}\ln \left[1-y(A+B)\right], \qquad K_1 = \frac{\tilde{K}_1(q=1)-\tilde{K}_1(q=-1)}{2} ,
\end{equation}
which generates all diagrams without fixing any of the channels to be the first.  This freedom allows an additional factor of $2n$ which can be obtained by differentiating to give the desired subleading order moment generating function
\begin{equation}
T_1 = r \frac{\rmd K_1}{\rmd r} .
\end{equation}

However, even when all of the tunneling probabilities are the same, where we can use \eref{phirecursummed} and \eref{phihatrecursummed} to simplify \eref{Beqn} and \eref{Aeqn}, since $\phi$ and $\ph$ are determined by quartic equations we were unable to obtain the corresponding quartic for $T_1$.  Of course we can substitute the expansions of $\phi$ and $\ph$ from \eref{phiexpansion} into \eref{Beqn} and \eref{Aeqn} and obtain arbitrarily many moments in the expansion of $T_1$:
\begin{eqnarray}
\fl T_1 &=& -\xi p s + 2 \xi p \left(1-2p-2\xi(3-4p)\right)s^2 \nonumber \\
\fl & & {} + \xi p \left(3p(2-3p)+\xi(18-93p+87p^2)-\xi^2(90-279p+205p^2)\right)s^3 + \ldots
\end{eqnarray}
The first two terms have been obtained previously \cite{whitney07,kuipers09,waltneretal12}, for the more general case of arbitrary tunneling probabilities which could likewise be incorporated in the expansions here.  The first term which is the weak localisation correction to the conductance agrees with the known semiclassical \cite{whitney07,kuipers09} and RMT \cite{bb96} result with equal probabilities, while the correction to the second moment matches the corresponding semiclassical result from considering the diagrams explicitly \cite{waltneretal12} which in turn is equal to the RMT diagrammatic result \cite{rbm08}. 

In the further simplified case of equal leads $\zeta_1=\zeta_2$ where $\phi=\ph$ are given by the quadratic in \eref{phiequalleads}, we can directly obtain the corresponding generating function
\begin{equation}
\label{transsubleading}
\fl T_1 = \frac{sp(1-p)(2-p)(2-s)\sqrt{1-s}-sp^2(1-p)(2-2s+s^2)-s^3p^4/4}{[4(1-p)+sp^2][4(1-p)(s-1)+sp^2](1-s)} .
\end{equation}

\end{document}